\documentclass[aps,prl,twocolumn,superscriptaddress]{revtex4-2}

\usepackage{ulem}
\usepackage{amsmath}
\usepackage{amssymb}
\usepackage{graphicx}
\usepackage{color}
\usepackage[colorlinks,citecolor=blue,linkcolor=red,urlcolor=blue]{hyperref}

\begin{document}
	
	\title{High Chern Number Quantum Anomalous Hall States in Haldane-Graphene Multilayers}
	
	\author{Yuejiu Zhao}
	\email{zhaoyuejiu@ucas.ac.cn}
	\affiliation{Kavli Institute for Theoretical Sciences, University of Chinese Academy of Sciences, Beijing 100190, China}
	
	\author{Long Zhang}
	\email{longzhang@ucas.ac.cn}
	\affiliation{Kavli Institute for Theoretical Sciences, University of Chinese Academy of Sciences, Beijing 100190, China}
	
	\author{Fu-Chun Zhang}
	\email{fuchun@ucas.ac.cn}
	\affiliation{Kavli Institute for Theoretical Sciences, University of Chinese Academy of Sciences, Beijing 100190, China}

	\date{\today}
	
	\begin{abstract}
		We consider a rhombohedral-stacked $N$-layer graphene coupled to a monolayer of Haldane model. We show that high order Dirac points in multilayer graphene can be gapped out by topological proximity effect of the Haldane model layer, leading to total Chern number $|C|=N+1$ quantum anomalous Hall states. This provides a new way to construct high Chern number quantum anomalous Hall states in realistic crystalline graphene systems.
	\end{abstract}		
	\maketitle
	
	\textit{Introduction.---}Quantum anomalous Hall (QAH) insulators \cite{Liu2016, Chang2023} exhibit quantized Hall conductance, $\sigma_{\mathrm{H}}=Ce^{2}/h$, without magnetic field, where $C$ is the total Chern number of the occupied bands \cite{Thouless1982}. The QAH effect was first proposed theoretically in the Haldane model on a honeycomb lattice \cite{Haldane1988}. In order to realize the QAH effect in laboratory, it is crucial to break the time reversal symmetry (TRS) and open an insulating gap in the bulk state. The QAH effect has been observed in two-dimensional (2D) materials with magnetic doping \cite{Chang2013} or intrinsic magnetism \cite{Gong2019}, where the TRS is broken by magnetic order while the energy gap is opened by spin-orbit coupling. In moir\'e superlattice materials \cite{Serlin2020}, the TRS is broken by the valley polarization while the gap is opened by electron interactions. 
	
	Based on the classification of topological band theory \cite{Chiu2016}, QAH systems are characterized with a integer $\mathbb{Z}$ index. High Chern number QAH states can be constructed theoretically by gapping out a parent Hamiltonian with multiple Dirac points at the Fermi energy \cite{Sticlet2012}. Examples include the Haldane-type model with long-range hopping \cite{Sticlet2013}, the double bilayer graphene with valley splitting and orbital magnetism \cite{WangY2021}, and twisted multilayer moir\'e structures \cite{WangH2024, Liu2019}. Another conceptually simpler approach is to stack multiple layers of $C=1$ QAH states together \cite{WangY2021a, WangD2023, Li2020,Liang2025}, which has been realized in the $\text{Cr}$-doped $(\text{Bi,Sb})_2\text{Te}_3$ \cite{Zhao2020} and the $\text{MnBi}_2\text{Te}_4$ multilayers \cite{Ge2020}.
	
	High Chern number QAH systems can also be constructed by gapping out high order Dirac points as well. The effects of high order Dirac points was firstly studied in gapless rhombohedral-stacked bilayer graphene system under a magnetic field \cite{McCann2006}, where an individual $2^{nd}$-ordered Dirac point carries a $2\pi$ Berry phase. When Coulomb interactions are considered, various symmetry breaking insulating ground states of graphene multilayers emerge \cite{Nandkishore2010, Zhang2011, Zhang2010, Jung2013}, some of which may host QAH or quantum spin Hall states \cite{Nandkishore2010, Zhang2011}. The high order Dirac points induced QAH states have been firstly detected in moir\'e superlattice \cite{Chen2020, WangS2024}, where the effects of electron interaction and spin-orbit coupling are enhanced due to the flat bands. Recently, high Chern number QAH states have been detected in crystalline graphene multilayers with spin-orbit proximitized \cite{Sha2024, Han2024}, where the band inversion are introduced by spin-orbit couplings and the topologically nontrivial ground states are found by fine-tuning the displacement field \cite{Liu2025}.
	
	In this Letter, a new approach to high Chern number QAH states by engineering and gapping out high-order Dirac points is proposed by topological proximity effects. We study the heterostructure formed by a rhombohedral-stacked $N$-layer graphene and a monolayer of Haldane model, and show numerically that the system hosts $|C|=N+1$ QAH phases. The high Chern number QAH state can be explained analytically by effective theory under two limiting conditions. In these limits, the effective two-band model of $N$-layer graphene has one pair of high order Dirac points, which is gapped out by the Haldane layer with opposite mass term, leading to $(N+1)^{\text{th}}$ order QAH states. The high Chern number QAH states in the heterostructure can be regarded as the ``topological proximity effect'' induced by the Haldane layer in the semimetallic graphene multilayer. 
	
\begin{figure}
		\includegraphics[width=0.5\textwidth]{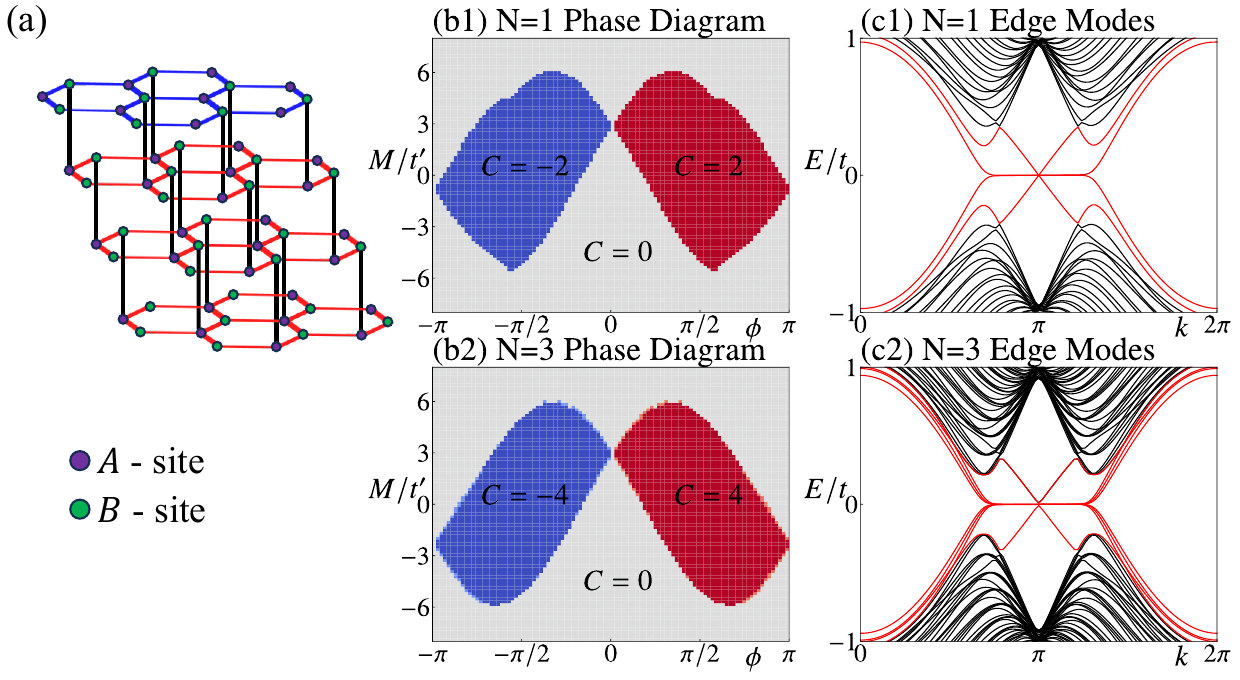}
		\caption{(a) Schematic of heterostructure composed of a layer of Haldane model \eqref{eq_Haldane} in blue color and rhombohedral-stacked $N=3$ layer graphene in red color.  The black vertical lines indicate interlayer hoppings. (b1) and (b2) show phase diagrams in parameter space $\left(\phi,M/t^\prime\right)$ for the total Chern number $C$ in \eqref{eq_Fukui} of the heterostructure \eqref{eq_Htotal} for $N=1$ and $N=3$ respectively. In the calculation, we choose $t=1$, $t^\prime=0.1$, $t_\perp=0.5$, and $t^\prime_\perp=0.1$. Phase diagrams are very similar for other values of $N$. (c1) and (c2) are the calculated dispersions with zigzag boundary in cylinder geometry. The red line shows edge modes. The stripe width is $L=10$ with $M=0$ and $\phi=\frac{\pi}{2}$. Other parameters are same as (b1) and (b2).}
		\label{fig_system}
	\end{figure}

	\textit{Haldane-graphene heterostructure.---} The high Chern number QAH states proposed in this Letter are formed by spinless electrons in an $(N+1)$-layer system, which contains one Haldane layer and $N$ graphene layers rhombohedral-stacked on each other, as shown in Figure \ref{fig_system} (a). The total Hamiltonian of this heterostructure is given by
	\begin{equation}
		\hat{H}=\sum_{i=0}^{N}\hat{H}_i+\sum_{i=1}^{N-1}\hat{H}^\perp_{i,i+1}+\hat{H}^\perp_\text{H}.
		\label{eq_Htotal}
	\end{equation}
	The top layer is described by the Haldane Hamiltonian on a honeycomb lattice \cite{Haldane1988},
	\begin{align}
		\hat{H}_0&=t\sum_{l,\delta}\left(c^\dagger_{0,l,A}c_{0,l+\delta,B}+h.c.\right)\notag\\&+M\sum_l\left(c^\dagger_{0,l,A}c_{0,l,A}-c^\dagger_{0,l,B}c_{0,l,B}\right)\notag\\&+t^\prime\sum_{l,\delta^\prime}\left(e^{i\phi}c^\dagger_{0,l+\delta^\prime,A}c_{0,l,A}+e^{-i\phi}c^\dagger_{0,l+\delta^\prime,B}c_{0,l,B}\right),
		\label{eq_Haldane}
	\end{align}
	where $t$ is the nearest-neighbor hopping amplitude, $A$ and $B$ are the sublattice indices, $l$ is the unit cell index, $\delta$ denotes the nearest-neighbor vectors and $\delta^\prime$ denotes the next-nearest-neighbor vectors. $\pm{M}$ is the chemical potential shift of the sublattices, and $t^\prime{e}^{\pm{i}\phi}$ are complex-valued hopping amplitudes, which explicitly break the TRS. The Hamiltonian of the $i^\text{th}$ graphene layer \cite{Neto2009} is given by
	\begin{equation}
		\hat{H}_i=t\sum_{l,\delta}c^\dagger_{i,l,A}c_{i,l+\delta,B}+h.c.,
	\end{equation}
	and $i=1,\dots,N$ are the layer indices. The nearest neighbour hopping amplitude is set the same as the Haldane layer for simplicity, whose difference would not violate the general results. The interlayer coupling of rhombohedral-stacked (or $ABC$-stacked in some literatures) multilayers is described by vertical hopping processes between $B$-atom sites of the $i^\text{th}$ graphene layer and $A$-atom sites on the $(i+1)^\text{th}$ graphene layer that are overlapped with each other as
	\begin{equation}
		\hat{H}^\perp_{i,i+1}=t_\perp\sum_{l}c^\dagger_{i,l,B}c_{i+1,l,A}+h.c.
	\end{equation}
	Couplings between the Haldane monolayer and the $N$-layer graphene is given by
	\begin{equation}
		\hat{H}^\perp_\text{H}=t^\prime_\perp\sum_{l}c^\dagger_{0,l,B}c_{1,l,A}+h.c.,
	\end{equation}
	and other interlayer processes are neglected. For simplicity, we assume $t_\perp^\prime\leq{t_\perp}$ in the following analysis.
	
	In the reciprocal space, the Hamiltonian is given by
	\begin{equation}
		\hat{H}=\sum_{k}\Psi^\dagger_kH_k\Psi_k,
	\end{equation}
	where the annihilation operator is arranged as a column vector $\Psi_k=\left(c_{0,k,A}{\quad}c_{0,k,B}{\quad}\cdots{\quad}c_{N,k,A}{\quad}c_{N,k,B}\right)^T$. The corresponding single-electron Hamiltonian is given by the following $2(N+1)\times2(N+1)$ tridiagonal matrix,
	\begin{equation}
		H_k=\left(\begin{array}{cccccccc}
			g^+_k    & f_k     & 0        & 0      & 0      & \cdots & 0      & 0\\
			f^\ast_k & g^-_k &t^\prime_\perp& 0      & 0      & \cdots & 0      & 0\\
			0        &t^\prime_\perp& 0        & f_k    & 0      & \cdots & 0      & 0\\
			0        & 0       & f^\ast_k & 0      & t_\perp& \cdots & 0      & 0\\
			0        & 0       &0         & t_\perp& 0      & \cdots & 0      & 0\\
			\vdots   & \vdots  & \vdots   & \vdots & \vdots & \ddots & \vdots & \vdots\\
			0        & 0       & 0        & 0      & 0      & \cdots & 0      & f_k\\
			0        & 0       & 0        & 0      & 0      & \cdots & f^\ast_k& 0\\
		\end{array}\right),
		\label{eq_Hk}
	\end{equation}
	where $g_k^\pm=\pm{M}+2t^\prime\sum_{\delta^\prime}\cos(k\cdot\delta^\prime\pm\phi)$ and $f_k=t\sum_{\delta}e^{ik\cdot\delta}$.
	
	The Hamiltonian \eqref{eq_Hk} can be solved numerically. Following Ref. \cite{Fukui2005}, the corresponding total Chern number of the occupied bands can be calculated by
	\begin{equation}
		C=\frac{1}{2\pi{i}}\sum_{n\in\text{occpd.}}\sum_{k\in\text{1BZ}}\ln\frac{U^n_x(k)U^n_y(k+\delta{k_x})}{U^n_y(k)U^n_x(k+\delta{k_y})}
		\label{eq_Fukui}
	\end{equation}
	with
	\begin{equation}
		U^n_{x/y}(k)=\frac{\langle{u^n(k)}|u^n(k+\delta{k_{x/y}})\rangle}{|\langle{u^n(k)}|u^n(k+\delta{k_{x/y}})\rangle|},
	\end{equation}
	where $|u^n(k)\rangle$ is the $n^\text{th}$ eigenvector of \eqref{eq_Hk}. High Chern number $C=\pm(N+1)$ QAH phases set in in the $(N+1)$-layer heterostructure, and the phase diagrams of $N=1,3$ are plotted in Fig. \ref{fig_system} (b), where QAH phases with total Chern number $C=\pm2$ and $C=\pm4$ are obtained. Because of the bulk-edge correspondence \cite{Hasan2010}, the Chern number of the bulk state equals to the number of chiral edge modes carrying dissipationless quantized Hall currents, which are shown in Fig. \ref{fig_system} (c). 
	
\begin{figure}
		\includegraphics[width=0.5\textwidth]{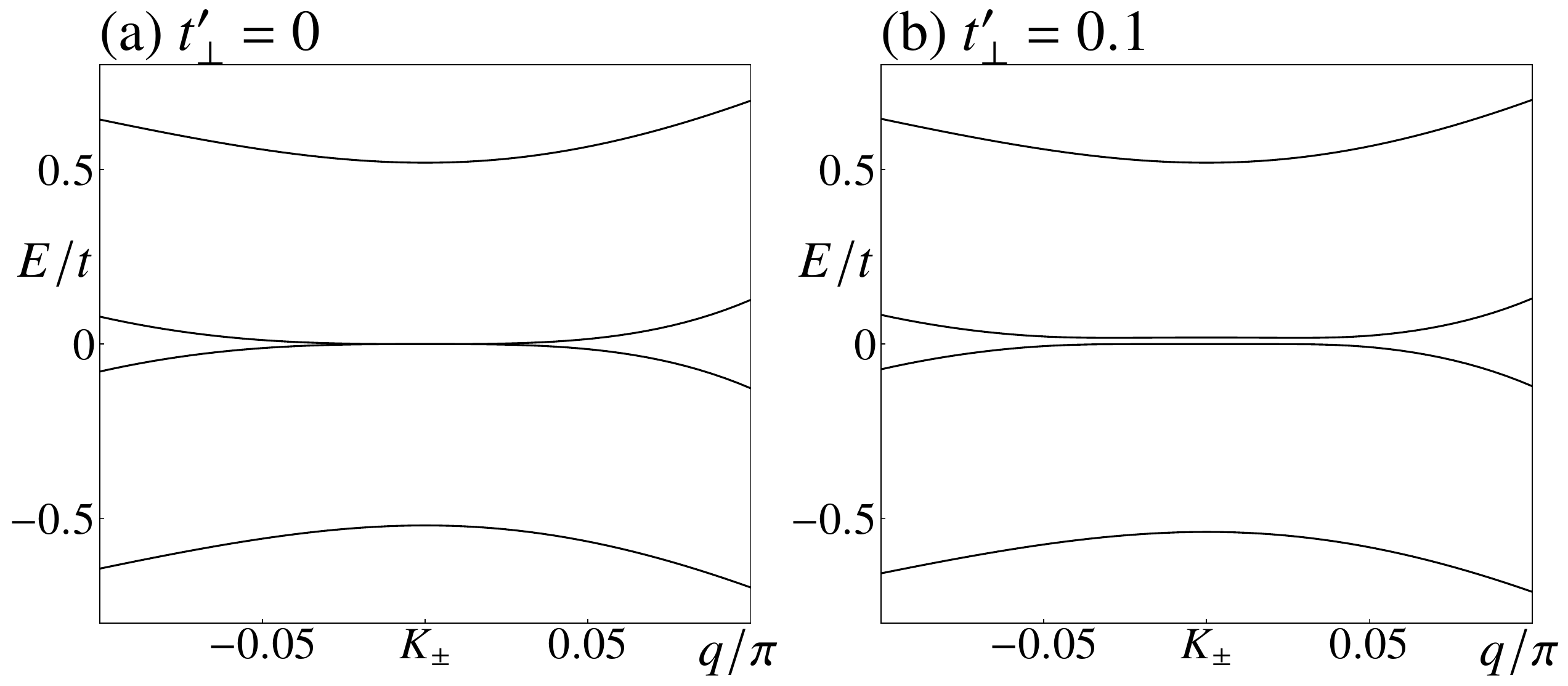}
		\caption{(a) shows the gapless dispersions $E(q)\sim\pm{q}^3$ near the Dirac points $K_\pm$ at $N=3$ when $t^\prime_\perp=0$. Other parameters are chosen as $t^\prime=0.1$, $t_\perp=1$, $M=0$ and $\phi=\pi/2$. (b) shows the degeneracy at Dirac points is lifted by the Haldane layer when $t^\prime_\perp$ is turned on.}
		\label{fig_band}
	\end{figure}

	While the Hamiltonian \eqref{eq_Hk} is not analytically solvable in general, it can be studied with effective models in certain limits. Firstly, we introduce the low-energy theory \cite{Min2008} of rhombohedral-stacked $N$-layer graphene under $(c_{1,k,A}\quad{c}_{N,k,B})$ basis as
	\begin{equation}
		H_N^{\text{grph}}(k)=\left(\begin{array}{cc}0&F_N(k)\\F^\ast_N(k)&0\end{array}\right),
		\label{eq_NLG}
	\end{equation}
	with $F_N(k)=(-t_\perp)^{1-N}f_k^N$. The low-energy degrees of freedom are hosted by $c_{1,k,A}$ and ${c}_{N,k,B}$ since they do not have interlayer coupling with other degrees of freedom, and form a gapless system effectively. Near the Dirac points of the monolayer graphene denoted by $K_\pm$, the dispersions regarding to $q=k-K_\pm$ are $E(q)\sim\pm{q}^N$, which means the Dirac points become $N^\text{th}$-order zero of the band dispersion. When the Haldane layer is stacked on, the effective Hamiltonian is 
	\begin{equation}
		H_\text{eff}(k)=\left(\begin{array}{cccc}g^+_k&f_k&0&0\\f^\ast_k&g^-_k&t^\prime_\perp&0\\0&t^\prime_\perp&0&F_N(k)\\0&0&F^\ast_N(k)&0\end{array}\right)
		\label{eq_Heff0}
	\end{equation}
	Dispersions of \eqref{eq_Heff0} near the Dirac points $K_\pm$ are shown in Figure. \ref{fig_band}. The system is gapless with $E(q)\sim\pm{q}^N$ dispersions without $t^\prime_\perp$. When an infinitesimal $t^\prime_\perp$ is turned on, the finite Haldane gap can gapped out the $N$-layer graphene and make the heterostructure an insulator, where the Chern number thus be well-defined. We have numerically checked that this Hamiltonian and \eqref{eq_Hk} give the same results of total Chern number, thus we are going to explain the $|C|=N+1$ phases regarding to \eqref{eq_Heff0}. The following of this Letter introduces the large Haldane gap limit, where the energy gap in the Haldane layer \eqref{eq_Haldane} is the leading energy scale; and the large $t^\prime_\perp$ limit, where the coupling between Haldane layer and graphene multilayer is the leading energy scale. The $C=\pm(N+1)$ QAH states can be obtained analytically in both limiting conditions, and the crossover between them does not destroy the high Chern number topological phases, thus the Haldane-graphene heterostructure can host high Chern number QAH states in a large parameter range.
	
	\textit{Large Haldane gap limit.---} When an infinitesimal $t_\perp^\prime$ is turned on, the leading energy scale in \eqref{eq_Heff0} is the initial Haldane gap. To ensure the system staying in this limit, we focus on $\phi=\pi/2$ in the following analysis. Within this regime, the Haldane layer can be regarded as not affected by stacking, while an effective gap will be induced into the graphene multilayers. The low-energy subspace near the Fermi surface is spanned by the $(c_{1,k,A}\quad{c}_{N,k,B})$ basis of graphene multilayer. The low-energy effective Hamiltonian can be derived from \eqref{eq_Heff0} via Schrieffer-Wolff transformation as
	\begin{align}
		&H_\text{grph}^\text{eff}(k)=H_\text{low}-TH_\text{high}^{-1}T^\dagger\notag\\
		=&\left(\begin{array}{cc}0&F_N(k)\\F^\ast_N(k)&0\end{array}\right)-\left(\begin{array}{cc}0&0\\t_\perp^\prime&0\end{array}\right)\left(\begin{array}{cc}g_k^+&f_k\\f_k^\ast&g_k^-\end{array}\right)^{-1}\left(\begin{array}{cc}0&t^\prime_\perp\\0&0\end{array}\right)\notag\\
		=&\left(\begin{array}{cc}0&F_N(k)\\F^\ast_N(k)&\frac{(t^\prime_\perp)^2{g_k^-}}{|f_k|^2-g_k^+g_k^-}\end{array}\right),
		\label{eq_Hgeff}
	\end{align}
	where $T$ and $T^\dagger$ are off-diagonal operators mapping into and out of the low-energy subspace, respectively.
	
	This effective Hamiltonian can be evaluated analytically in the continuum limit near each Dirac point. In the polar coordinates, $q_x=q\cos\theta$ and $q_y=q\sin\theta$, \eqref{eq_Hgeff} can be written as $H_\text{grph}^\text{eff}(q)=\alpha\vec{d}_\xi(q,\theta)\cdot\vec{\sigma}$ up to a constant, where $\alpha$ is a coefficient related to the Fermi velocity at $K_\xi$. The corresponding $d$-vector is $\vec{d}_\xi(q,\theta)=(\xi{q}^{N}\cos(N\theta),q^{N}\sin(N\theta),M_\xi)$ with $M_\xi=(t_\perp)^{N-1}(t^\prime_\perp)^2/2(M+\xi3\sqrt{3}t^\prime)$, thus Berry curvature is calculated as
	\begin{equation}
		\Omega_\xi(q,\theta)=\frac{\vec{d}\cdot(\partial_{q_x}\vec{d}\times\partial_{q_y}\vec{d})}{2|\vec{d}|^3}=\frac{\xi{M_\xi}N^2q^{2N-2}}{2(q^{2N}+M_\xi^2)^{3/2}}.
	\end{equation}
	The integration gives that the Chern number of each Dirac point is
	\begin{equation}
		C^g_\xi=\frac{1}{2\pi}\int_{0}^{2\pi}\text{d}\theta\int_{0}^{\infty}q\Omega_\xi(q,\theta)\text{d}q=\xi\frac{N}{2}\text{sgn}M_\xi.
		\label{eq_C}
	\end{equation}
	(The Chern number can be calculated by alternative methods like \cite{Cheng2022,Andrijauskas2015}). Therefore, \eqref{eq_Hgeff} can host $C^{g}=N$ phases if
	\begin{equation}
		|M|<3\sqrt{3}t^\prime.
		\label{eq_GapPB}
	\end{equation}
	Together with the Haldane monolayer, whose phase boundary of $C^{H}=1$ states at $\phi=\pi/2$ is given by \eqref{eq_GapPB} as well, the heterostructure can host $C=N+1$ QAH states at the large Haldane gap limit. Similar analysis works for $\phi=-\pi/2$ case, where the heterostructure can host $C=-(N+1)$ QAH states. It is worth noting that an infinitesimal $t^\prime_\perp$ can gap out the graphene multilayer and make the heterostructure a $|C|=N+1$ insulator.
	
\begin{figure}
		\includegraphics[width=0.5\textwidth]{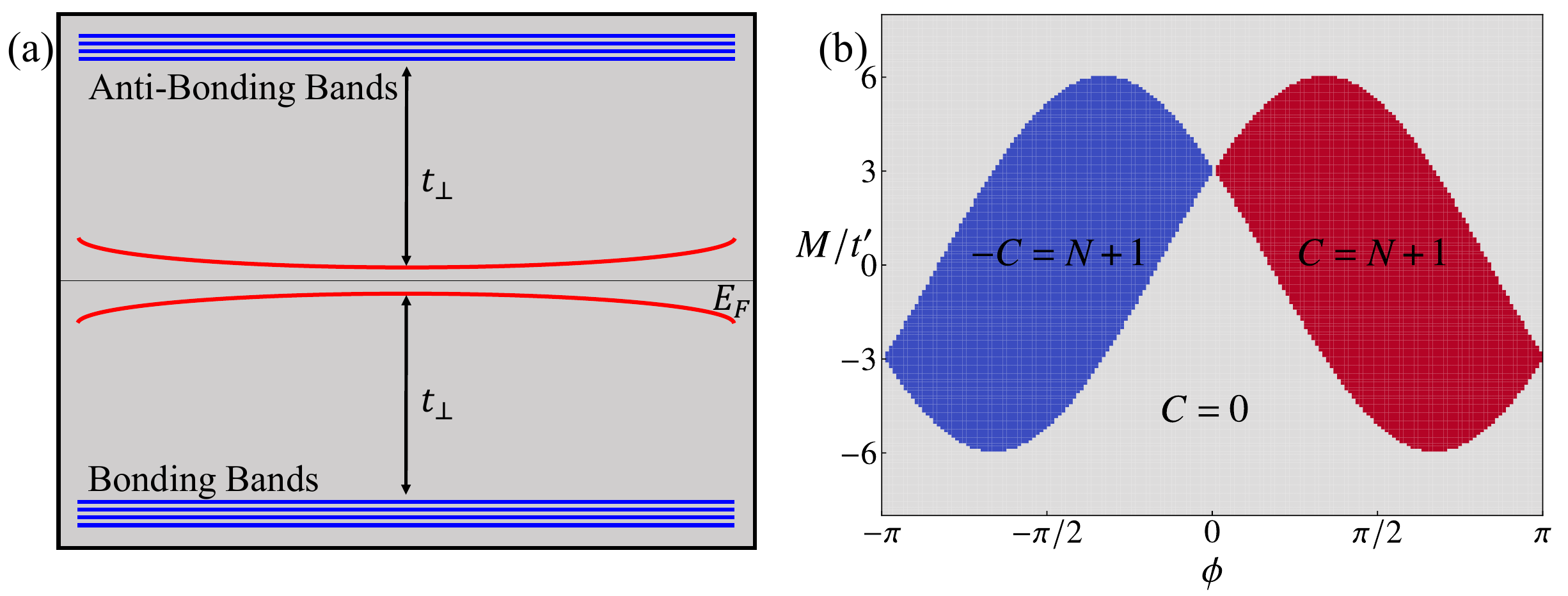}
		\caption{(a) illustrates the band structure in the large $t^\prime_\perp$ limit. The blue lines indicates the low(high)-energy bands formed by (anti-)bonding states. The effective theory are based on the red bands near the Fermi surface whose degrees of freedom come from $c_{0,k,A}$ and $c_{N,k,B}$. (b) shows the phase diagram of \eqref{eq_Htv}. The topological phase boundaries are given by \eqref{eq_boundary}, where the $C=\pm(N+1)$ phases are meet at $\phi=\pm\pi$, $M=-3$ and $\phi=0$, $M=3$.}
		\label{fig_largeT}
	\end{figure}

	\textit{Large $t^\prime_\perp$ limit \& crossover.---} When $t^\prime_\perp$ is increased to be the leading energy scale of \eqref{eq_Heff0}, we study the heterostructure in the large $t^\prime_\perp$ limit. For simplicity, we assume $t^\prime_\perp=t_\perp$ in the following analysis. In this limit, the hopping amplitudes between the single-electron states belonging to the $i^\text{th}$ layer and the $(i+1)^\text{th}$ layer are large, thus these states will hybridize to form bonding and anti-bonding states, whose characteristic energies are $\mp{t^\text{(H)}_\perp}$. Therefore, the effective theory near the Fermi surface is formed by single-electron states of the $A$-sites in the Haldane layer and the $B$-sites in the $N^\text{th}$ graphene layer, which is illustrated in Fig. \ref{fig_largeT} (a). Since the bonding and anti-bonding states are highly localized, the topological properties of the multilayer system are captured by the valence band near the Fermi surface. Due to the large gap between the rest bands and the low-lying occupied bands formed by bonding states, summation of their Chern number should be zero even though they might hybridize with each other and make it possible that a single band can have nonzero Chern number.
	
	In this limit, the effective Hamiltonian matrix under $(c_{0,k,A}\quad{c}_{N,k,B})$ basis can be derived by a Schrieffer-Wolff transformation from \eqref{eq_Heff0} similarly to \eqref{eq_Hgeff} as
	\begin{equation}
		H^{t_\perp^\prime}(k)=\left(\begin{array}{cc}g^+_k&F_{N+1}(k)\\F^\ast_{N+1}(k)&0\end{array}\right),
		\label{eq_Htv}
	\end{equation}
	and the virtual processes between $B$-sites of the graphene layer are neglected since they are higher order perturbations. The Haldane layer breaks the TRS and introduces a mass term to make the multilayer an insulator.
	
	The effective Hamiltonian matrix \eqref{eq_Htv} can be written in a canonical form as $H^{t_\perp^\prime}(k)=\sum_{i=0}^3d_N^i(k)\sigma^i$, where $\sigma^i$s are Pauli matrices acting on sublattice space. The $d$-vector is determined by $d_N^1(k)=\text{Re}F_{N+1}(k)$, $d_N^2(k)=-\text{Im}F_{N+1}(k)$, and $d_N^0(k)=d_N^3(k)=g^+_k/2$. The gap closing conditions, which correspond to possible topological phase boundaries, are determined by $d_N^1(k)=d_N^2(k)=d_N^3(k)=0$. Solutions of $d_N^1(k)=d_N^2(k)=0$ are $K_\xi$ with $\xi=\pm$, under which the gap closing condition $d_N^3(K_\xi)=0$ gives
	\begin{equation}
		6\sin\left(\frac{\pi}{6}+\xi\phi\right)-\frac{M}{t^\prime}=0
		\label{eq_boundary}
	\end{equation}
	as possible topological phase boundaries. The Chern number of the valence band of \eqref{eq_Htv}, or the total Chern number of \eqref{eq_Hk}, within each region in the phase diagram can be calculated similarly to \eqref{eq_C}. The total Chern number of each region in the phase diagram is the summation of two Dirac points', as shown in Figure \ref{fig_largeT} (b). This phase diagram is different to the Haldane model's \cite{Haldane1988} since the topological gap is opened by the next nearest-neighbour hoppings within $A$-sublattice only. This result reveals that the $N$-layer graphene together with one Haldane layer can host $|C|=N+1$ QAH phases since the Dirac points $K_\pm$ become $(N+1)^\text{th}$-order zeros of $d_1^N$ and $d_2^N$ due to interlayer virtual processes. 
	
\begin{figure}
		\includegraphics[width=0.5\textwidth]{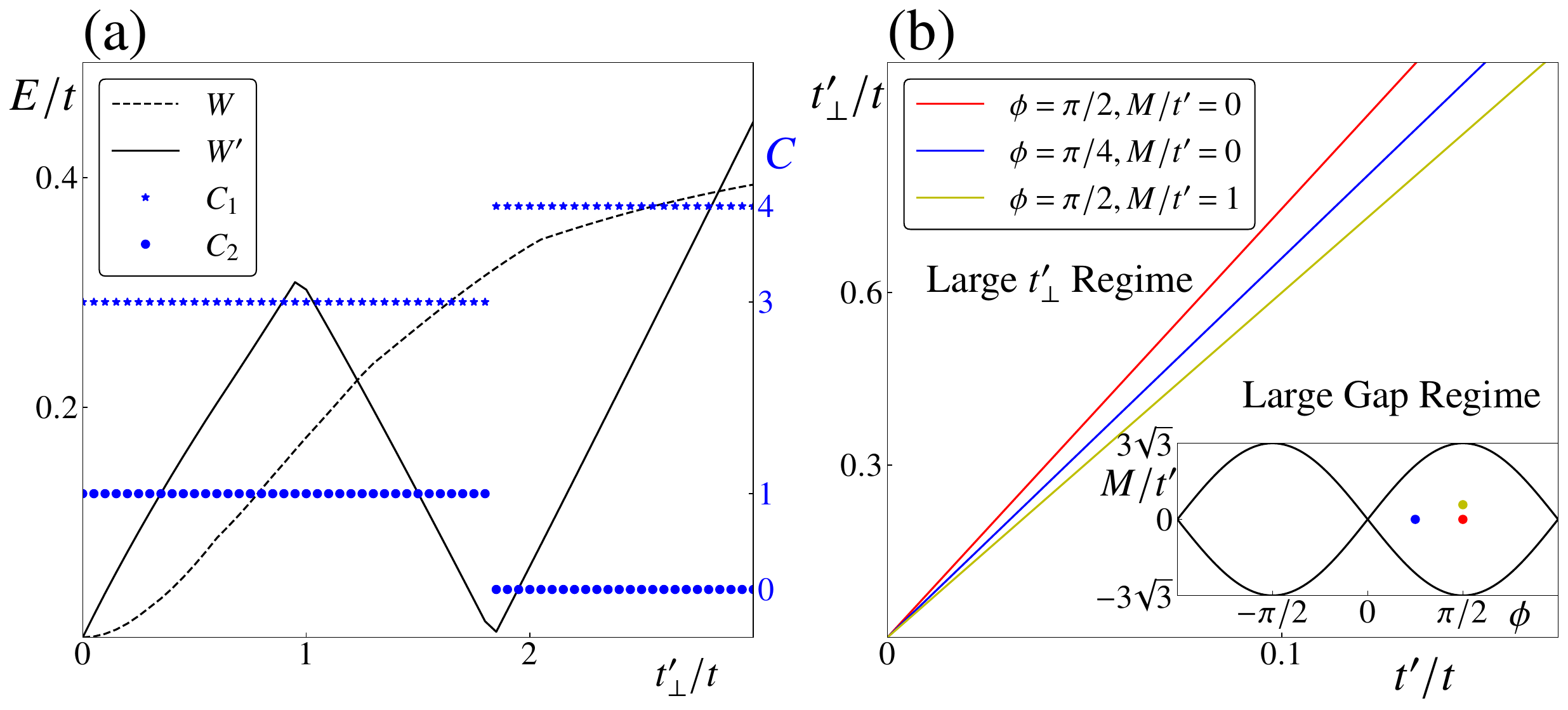}
		\caption{(a) shows the exchange of Chern number between two occupied bands with $t^\text{H}_\perp$ increasing at $N=3$. Other parameters are set as $t_\perp=1$, $t^\prime=0.1$, $M=0$ and $\phi=\frac{\pi}{2}$. The dashed line shows the gap $W$ between occupied and vacant bands, which verifies the system being an insulator at finite $t^\prime_\perp$. The full line shows that the gap $W^\prime$ between two occupied bands closes, at which the exchange of Chern number happens, as shown by the blue stars and dots corresponding to the Chern numbers of the upper and lower occupied bands. (b) shows the phase diagram of \eqref{eq_Heff0} at $N=3$. The total Chern number is always being $C=4$ and boundaries of two limit regimes regarding to different Haldane model parameters are labeled by colors, which are denoted in the phase diagram of Haldane monolayer as shown in the inner figure.}
		\label{fig_cross}
	\end{figure}

	We have shown that the $|C|=N+1$ QAH phases can be understood by two limit regimes. In the large Haldane gap limit, the system has two topological occupied bands, one has $|C_1|=N$ and the other has $|C_2|=1$; in the large $t^\prime_\perp$ limit, the system has one topological occupied band with $|C|=N+1$. The crossover between these regimes can be understood via the Chern number exchange \cite{Guan2024} between two occupied bands. By numerically diagonalizing \eqref{eq_Heff0} at $N=3$, this exchange is shown in Figure \ref{fig_cross} (a). Since the gap at the Fermi surface keeps opening, the bilayer system keeps being an insulator, and the total Chern number is invariant. However, once the gap between two occupied bands close when $t^\prime_\perp$ keeps increasing, the Chern number exchange happens, and the system transforms from the large Haldane gap limit to the large $t^\prime_\perp$ limit. By varying $\phi$ from $\pi/2$ or introducing finite $M$, the Haldane gap is effectively reduced, and the boundaries of crossover is changed, as shown in Figure. \ref{fig_cross} (b).
	
	\textit{Summary \& discussions.---} In this Letter, a multilayer heterostructure that can host $|C|=N+1$ QAH states is proposed by rhombohedral stacking of $N$ graphene layers and one Haldane layer. The QAH states can be regarded as the results of topological proximity effects induced by the Haldane layer, which breaks the TRS and gapping out the high order Dirac points. This is a new regime of topological proximity effects that is different from the surface \cite{Zhang2014,Shoman2015} or bulk \cite{Hsieh2016,Cheng2019} ones since it can reserve low-energy degrees of freedom and keep the whole system topologically nontrivial even at the large coupling limit.
	
	The topological proximity effect may pave a new way to construct crystalline graphene multilayers with high Chern number QAH states in experiments. Recently, the Haldane model has been realized in moir\'e materials \cite{Zhao2024}. Since the high Chern number QAH states are governed by the low-energy physics near the Dirac points, the topological proximity effects induced gap cannot be violated by superlattice effects, which means the high Chern number states in rhombohedral-stacked graphene multilayers can be constructed by a nearby Chern insulator with different lattice constants. It is worth to note that the electron interactions might be crucial in systems with high-order Dirac points by introducing an insulating gap when the interacting strength is finite. However, if the proximity effect induced gap is larger than the interacting one, the system can host high Chern number QAH states with the topological phase boundaries modified. For realistic materials, such comparisons depend on detailed calculations based on realistic parameters, which is left for further studies.
	
	\begin{acknowledgements}
		This work is supported by China Ministry of Science and Technology (Grant No. 2022YFA1403902), Chinese Academy of Sciences (Grant Nos. YSBR-057 and JZHKYPT-2021-08), the National Natural Science Foundation of China (Grant No. 11804337), and the Innovative Program for Quantum Science and Technology (Grant Nos. 2021ZD0302500 and 2021ZD0302600). 
	\end{acknowledgements}
	
	\bibliography{refer}
\end{document}